\documentclass[aps,pre,twocolumn,superscriptaddress,floatfix]{revtex4}
	
\usepackage{graphicx}
\usepackage{epsfig}

\begin{document}

\title{$H-T$ phase diagram of the bi-dimensional Ising model with exchange and dipolar interactions}

\author{Rogelio D\'\i az-M\'endez}
\affiliation{Nanophysics Group, Department of Physics,
Electric Engineering Faculty, \\
CUJAE, ave 114 final, La Habana, Cuba}
\affiliation{``Henri-Poincar\'e-Group'' of Complex Systems, Physics Faculty, University of
Havana, La Habana, CP 10400, Cuba}

\author{Roberto Mulet}
\affiliation{``Henri-Poincar\'e-Group'' of Complex Systems, Physics Faculty, University of
Havana, La Habana, CP 10400, Cuba}
\affiliation{Department of Theoretical Physics, Physics Faculty, University of
Havana, La Habana, CP 10400, Cuba}

\date{\today}

\begin{abstract}

We explore the equilibrium properties of a two-dimensional Ising spin model with short-range exchange and long-range dipolar interactions as a function of the applied magnetic field $H$. The model is studied through extensive Monte Carlo simulations that show the existence of many modulated phases with long range orientational order for a wide range of fields. These phases are characterized by different wave vectors that change discontinuously with the magnetic field.
We provide numerical evidence supporting the existence of first order transitions between these phases. At higher fields our results suggest a Kosterliz-Thouless scenario for the transition from a bubble to a ferromagnetic phase.

\end{abstract}

\pacs{}

\maketitle

\section{Introduction}

Thin magnetic films have been the subject of intense attention over the last two decades \cite{mac95,wu05,vater00}.  Most studies have been motivated mainly by the technological applications of these structures \cite{bader06}. But, they also faced statistical physicists with the challenge of trying to answer many foundational questions regarding the role of microscopic interactions  in the macroscopic behavior of a large system. 

These quasi two-dimensional structures show a large variety of ordering effects including formation of striped states, reorientation transitions, bubbles formation in presence of magnetic fields and hysteresis \cite{all92,kash93,mar07}. At the origins of these phenomena is the competition between a short-ranged interaction favoring local order and a long-range interaction frustrating it on larger spatial scales. The role of the long-range interaction is to avoid the {\em global } phase separation favored by the short-ranged interaction promoting, instead, a state of phase separation at {\em mesoscopic or nano-scales}. Then, it is not, in general, a small perturbation,\cite{barci07} but must be considered as precisely as possible.

From a computational point of view, this means that the frustrating interaction has to be accounted for by involving all the lattice sites in the computation, which in turn limits the actual system size that can be handled in Monte Carlo simulations. On the other hand, to obtain exact results on multi-scale, multi-interaction systems is extremely difficult, so that simulations are often the only source of information. 

Model Hamiltonians taking into account short-ranged exchange  ferromagnetic  and long-range dipolar anti-ferromagnetic interactions have been used to reproduce many of the elemental features observed in experiments of magnetic systems \cite{hub98}. Unfortunately, and despite the obvious relevance from the experimental point of view of the presence of an external magnetic field, most 
of the numerical studies so far have concentrated their attention on the zero magnetic field case ($H=0$). This is in part because of the already very rich and complex phenomenology obtained by tuning the strengths of the exchange and the dipolar interactions, but also because of the almost prohibitive computational cost of the simulations, even for moderated lattice sizes. 
 
To fill this gap, we use extensive Monte Carlo simulations to determine the role of an external magnetic field in the thermodynamical properties of quasi two-dimensional magnetic systems. We present results for systems where exchange and dipolar interactions are comparable and where the anisotropy contribution to the Hamiltonian is very large.

The work is organized as follows: In section \ref{model} we present the model and review some of its  properties. In section \ref{sim} we give details about the Monte Carlo simulations and discuss the motivation for the parameters used and its connections with previous reports in the literature. Then, in section \ref{rd} we present and discuss our results. This section is organized in three parts, we first present and analyze the $H-T$ phase diagram of the model, then we provide some insight on the ground state structure of the different phases, and finally we characterize the transitions between these phases. Finally, in section \ref{conc} the conclusions of the work appear.

\section{The Model}
\label{model}

We consider a square lattice of Ising spins oriented perpendicularly to the plane of the lattice and interacting through the dimensionless Hamiltonian

\begin{equation}
 {\cal H} = -\delta\sum_{\langle ij\rangle }S_iS_j+
\sum_{i\neq j}\frac{S_iS_j}{r_{ij}^3}-H\sum_i S_i
\label{ham}
\end{equation}

\noindent where $S_i=\pm1$ is the value of the  spin at site $i$.  The first sum runs over all pairs of nearest neighbor spins and the second over all pair of spins in the lattice. The discreteness of $S_i$ is consistent with infinite or very large magnetic anisotropy \cite{hub98}. The parameter $\delta=J_e/J_d$ stands for the ratio between the strength of the exchange and dipolar interactions, $J_e$ and $J_d$ respectively. $H$ is the magnetic field intensity (in units of $J_d$) and $r_{ij}$ is the distance, measured in crystal units, between sites $i$ and $j$.

This model, but in zero external magnetic field, has been extensively studied\cite{deb00,cannas06,booth95,cannas04,arlett96}. For example, it is now well understood that in a wide range of values
of $\delta$ its ground state consists in stripes of anti-parallel spins with a width that increases with $\delta$ \cite{booth95}. Once the temperature is turned on, the situation becomes more complex and, in a $\delta-T$ phase diagram, one can recognize a zoology of phases, stripes of different widths, paramagnetic phases, tetragonal, smectic, nematic, and others \cite{mac95,cannas04}.  Roughly speaking, at zero field the system presents a first order phase transition between a low temperature phase of stripes and a high temperature tetragonal phase with broken translational and rotational symmetry. It was also shown \cite{cannas06} that for a narrow window around $\delta=4$ the model develops a nematic phase where the system has short range positional order but long range orientational order.

On the other hand, in ref. \cite{garel82} Garel and Doniach study analytically the $H-T$ phase diagram of a continuous Landau-like model with dipolar interactions. They conclude that the $H-T$ plane is 
characterized by three different phases: stripes, bubbles and ferromagnetic. Their analysis also suggests a 
scenario with Fluctuation Induced First Order Transitions (FIFOT)\cite{bra75} between the phases, or a second-order melting 
of the Kosterlitz-Thouless (KT)\cite{kost73} type for the bubble-ferromagnetic transition. 
While some of this phenomenology is confirmed by our simulations, we will show below that the phase diagram resulting from the
Hamiltonian (\ref{ham}) is even richer.

Numerical simulations using Langevin Dynamics\cite{fer06,jagla04,nicolao07} on  similar Landau-like models seem to support the general picture described in \cite{garel82}. In particular, in ref. \cite{jagla04} the author studies the behavior of the system under external magnetic field, but focus his attention mainly on the role of metastable configurations, the presence of hysteresis loops and memory effects. Therefore, the predictions of ref. \cite{garel82} are still waiting for conclusive numerical support.

For the particular case of the Hamiltonian (\ref{ham}), the correctness of the predictions of Garel and Doniach \cite{garel82} is even less clear. While at first one expects that the correspondence between the standard ferromagnetic Ising model and the continuous $\phi^4$ model persists even in the presence of the dipolar term, the existence of commensuration effects, typical of striped patterns in discrete Ising systems, may alter this intuition. For example, the authors of reference \cite{arlett96} studied  
the $H-T$ phase diagram of Hamiltonian (\ref{ham}) using  Montecarlo simulations
and found no evidence for the transition to a bubble phase, suggested a continuous character for a stripe-tetragonal boundary and reported some unexpected jumps in the  magnetization versus temperature curves.

In this sense our work revisits these previous simulations looking with more attention to the effect of the magnetic field at low temperatures. Some of the results already seen in \cite{arlett96} are confirmed and, we think, analyzed in more detail and from a different perspective. Some results support the predictions of \cite{garel82}, and others, to our knowledge, are new, and enrich the already complex phenomenology of these systems.

\section{Simulation}
\label{sim}

We centered our analysis on the value $\delta=4$ which corresponds to a zero field ground state of perfect alternating stripes of width $h=2$. So, for the smaller system sizes considered we have $8$ periods of modulated stripes. In all cases the size of the system $L$ was properly commensurate with the period of the $H=0$ modulated phase. Arlett et al. \cite{arlett96} used values of $\delta$ between $6$ and $8$, having stripes of width $h=4$ and $6$ respectively. So, for the system sizes they consider $4$ or at most $6$ periods of the modulated structures are present. As we will discuss below this makes difficult to interpret some of the consequences of the presence of $H$. 

This value of  $\delta=4$ is representative for proved first order stripes-tetragonal transition in $H=0$ but it is also known to be of the order of real magnetic-frustrated systems seen in experimental works \cite{deb00}. 
Some connections between experimental systems and values of relative strengths of interactions in theoretical models can be found in reference \cite{bland95}.  More recently, Carubelli et al. \cite{mar07} qualitatively reproduced detailed measurements of magnetic changes of samples of Fe/Ni/Cu(001) \cite{wong05} by means of a Heisenberg-spins model, very similar to Hamiltonian (\ref{ham}), using a value of $\delta=6$.

To build the phase diagram, the system is first initialized in the equilibrium configuration at a fixed temperature and zero magnetic field. To guarantee equilibration the magnetic field is increased very slowly
$10^{-4}\leq\Delta H\leq10^{-2}$ and for each $(H,T)$ point, 
we let the system relax for $t_1=10^6$ Montecarlo steps (mcs) using a Metropolis dynamics. Once equilibrated, the system evolves over other $t_2=10^7$ mcs to measure the physical quantities of interest. We impose periodic boundary conditions to limit finite size effects
and explore different values of temperature and field for linear system sizes up to $L=48$.  To account for the long-range interactions we implement the Ewald Summation Technique \cite{all94} adapted to the particular case of the magnetic dipolar potential \cite{rdm03}.

\section{Results and Discussion}
\label{rd}

\subsection{Phase Diagram}

\begin{figure}[!htb]
\includegraphics[width=6cm,height=7cm,angle=-90]{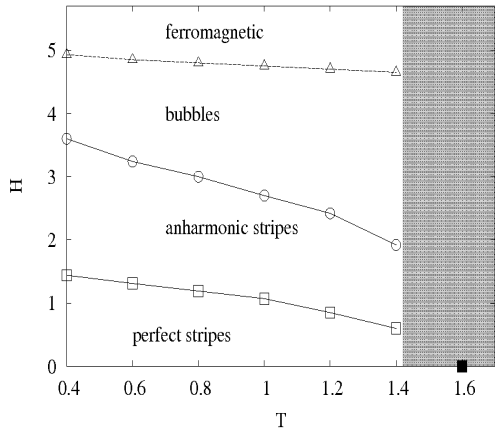}
\caption{Phases diagram for a system of L=32 considering the anharmonic zone; void circles and squares transition lines are first order. The critical temperature for the transition to the tetragonal phase at $H=0$ is shown with a colored square, the high temperature zone is represented with a slight shadow.}
\label{diag}
\end{figure}

The main result of this work is shown in figure \ref{diag}. 
This is the $H-T$ phase diagram of the model represented by the Hamiltonian (\ref{ham}).  

Four different zones are well defined in the diagram. For low values of temperature and external magnetic field the system is in a oriented  modulated phase of perfect stripes characterized by a wave vector $\vec{k}=(0,\pi/2)$ and zero magnetization. Increasing the magnetic field, new modulated phases, characterized by new wave vectors, and non-zero magnetization appear. These new phases, keep the orientational order but are characterized by several wave vectors (therefore we call them {\em anharmonic phases}) that depend on the magnetic field. The  properties of these phases and the location of the transitions suffer from strong finite size and commensuration effects, so, in the diagram we represented only one zone that, for the system size considered, contains all the anharmonic structures. Similar phases were already predicted within a mean field scenario for an Ising model with competing interactions $J_0$ and $J_1$ between nearest and next nearest neighbours in one direction of a cubic lattice ({\em ANNNI} model) \cite{Yokoi81}.

For still larger values of  $H$  we find a phase without orientational order ({\em bubble}). Finally, increasing further the magnetic field
  the system becomes completely magnetized (ferromagnetic phase).
At low $H$, and close to the stripe to tetragonal transition the combination between thermal fluctuations, commensuration and finite size effects, and the excitations due to the magnetic field makes the analysis of the phase diagram too difficult. So, in this zone, the structure of the phase diagram is still unknown, and we shadow this zone in figure \ref{diag} to caution the reader about this.
 
Now, to fix the ideas, let us concentrate our attention on the results for one temperature. We define, following \cite{booth95}, the so called $\pi/2$ rotational symmetry-breaking (SB) parameter:

\begin{equation}
\eta=|\frac{n_v-n_h}{n_v+n_h}|
\end{equation}
where $n_v$ ($n_h$) is the number of vertical (horizontal) bonds between nearest neighbors anti-aligned spins. This parameter takes the value $1$ in a perfectly ordered stripe state while it equals zero for any phase with $\pi/2$ rotational symmetry.

\begin{figure}[!htb]
\includegraphics[width=7cm,height=7cm]{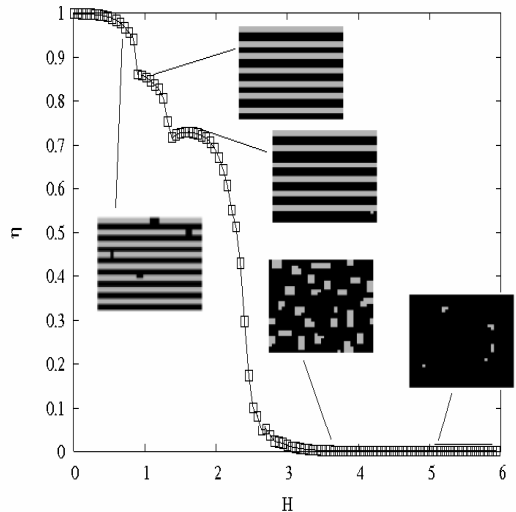}
\caption{SB parameter in a system of $L=32$ as a function of the field for $T=1.2$, the spots are some typical configurations.}
\label{pconf}
\end{figure}

In figure \ref{pconf} we represent the evolution of $\eta$ as a function of $H$ for $T=1.2$ in a system with $N=32 \times 32$ spins. The zooms  show typical configurations for the corresponding values of $\eta$. As can be seen, abrupt jumps separate clear plateaus of $\eta$ at three different values of the magnetic field, $H\simeq 0.84$, $H \simeq 1.34$ and $H \simeq 2.40$. 
Each plateau reflects an underlying symmetry of the system.

\begin{figure}[!htb]
\includegraphics[width=5.2cm,height=6.0cm,angle=-90]{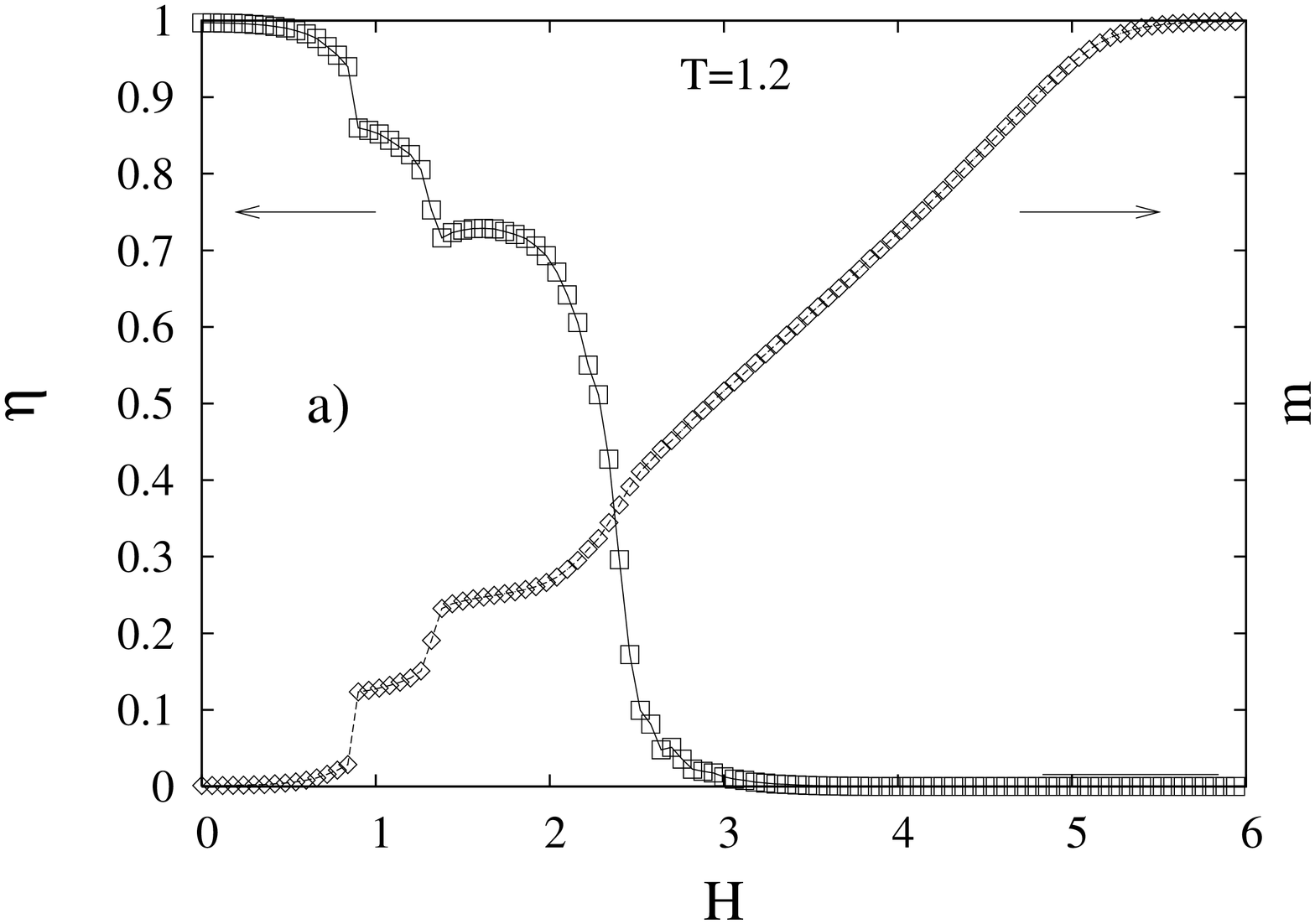}\\
\includegraphics[width=5.2cm,height=6.4cm,angle=-90]{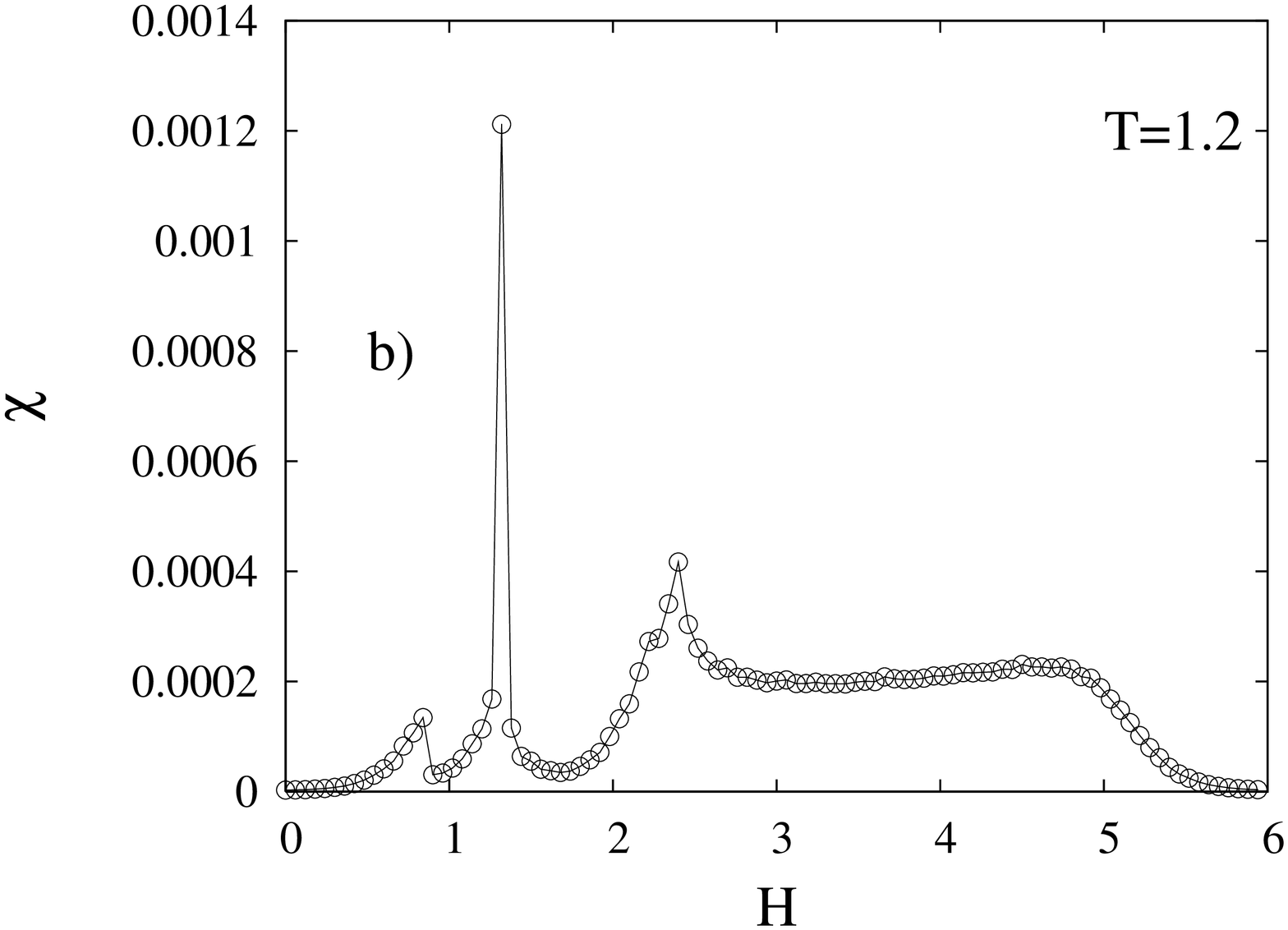}\\
\includegraphics[width=5.2cm,height=6.0cm,angle=-90]{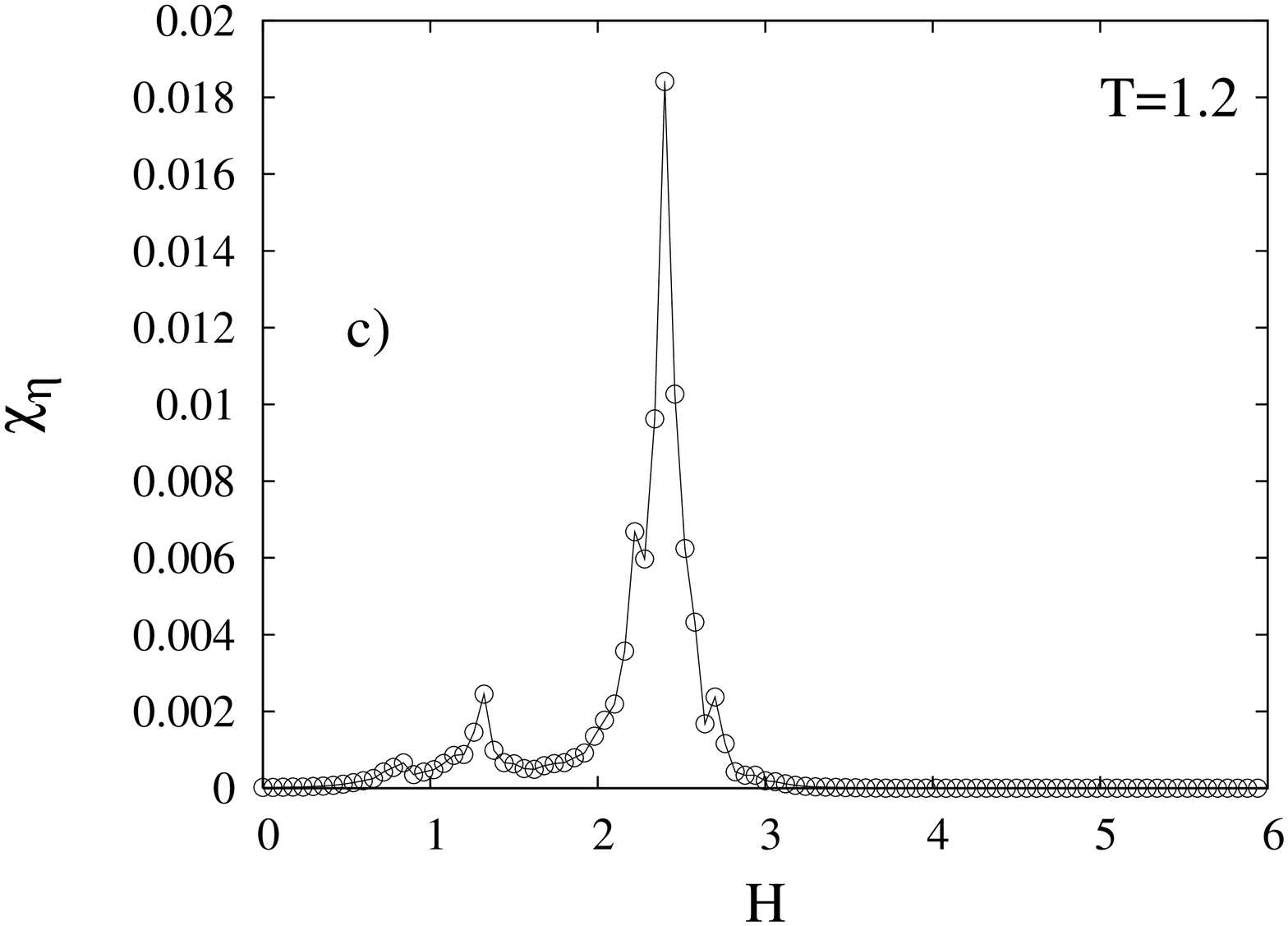}\\
\caption{Evolution under increasing magnetic field for a system of L=32: a) SB parameter and magnetization for $T=1.2$, b) magnetic susceptibility for $T=1.2$ and c) SB parameter associated susceptibility for $T=1.2$.}
\label{evo}
\end{figure}

A deeper understanding of the phase diagram and specially on the character of the jumps separating the different plateaus is obtained analyzing figure \ref{evo} where 
the magnetization, the magnetic susceptibility and the susceptibility associated to the rotational order parameter $\eta$ ($2 T \chi_{\eta}=<\eta^2>-<\eta>^2$) are plotted. Increasing from zero the external magnetic field,  the rotational symmetry-breaking  parameter $\eta$ and the magnetization show various plateaus separated by abrupt jumps (see fig. \ref{evo}a). These jumps result from the discrete properties of the lattice where the model is defined. Discrete 
changes in the field are required to change from one stable structure of stripes to another.

 The existence of these jumps is
also clearly reflected in both susceptibilities (see figures \ref{evo}b and \ref{evo}c). Three different peaks are well defined  in the magnetic and the orientational susceptibility at the same transition points where 
the orientational order parameter and the magnetization jump. 

For $H>3$,
the magnetization starts to growth linearly with $H$ but the rotational symmetry-breaking  parameter is zero. The system is in the so called bubble phase already predicted by Garel and Doniach \cite{garel82} for the Ginzburg-Landau model with dipolar interaction. Finally at very  high fields ($H>5$) the system is completely magnetized, $m=1$ and $\eta=0$. 

The jumps in the order parameter and the peaks in the susceptibilities suggest the existence of different thermodynamic phases at each plateau of $\eta$. To characterize the properties of these phases we look at the form of the structure factor, $S(\vec{k})$, in each plateau :

\begin{equation}
S(\vec{k})=\langle\mid \sum_iS_ie^{-i\vec{k}\cdot\vec{r_i}}\mid^2\rangle
\end{equation}

\begin{figure}[!htb]
\includegraphics[width=4.2cm,height=4.2cm]{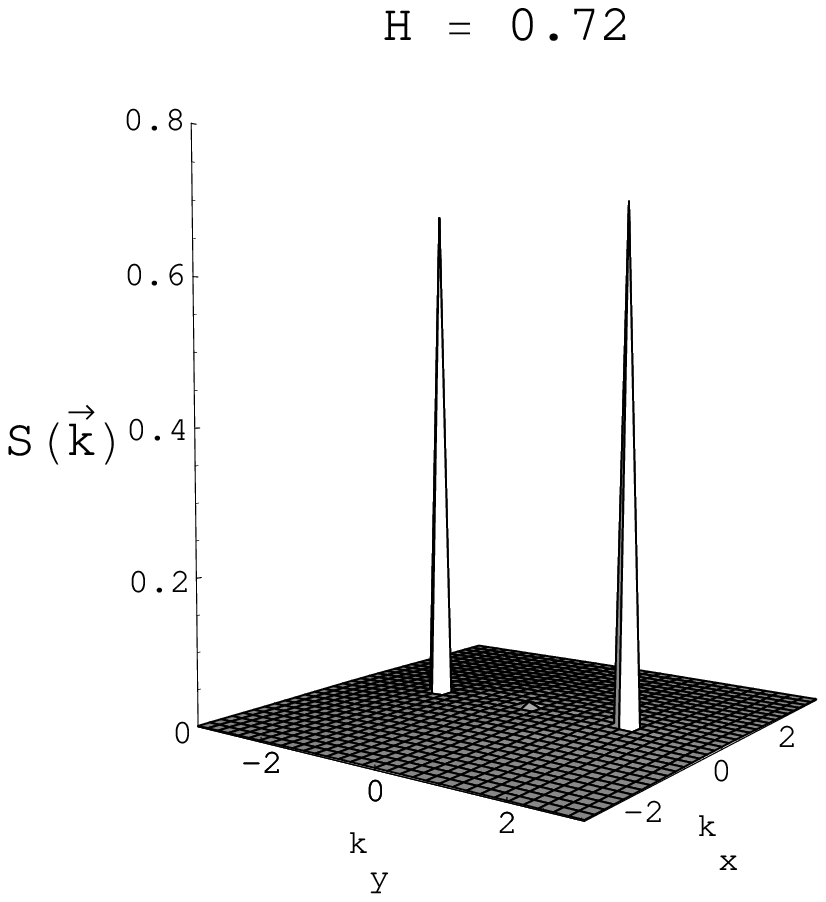}
\includegraphics[width=4.2cm,height=4.2cm]{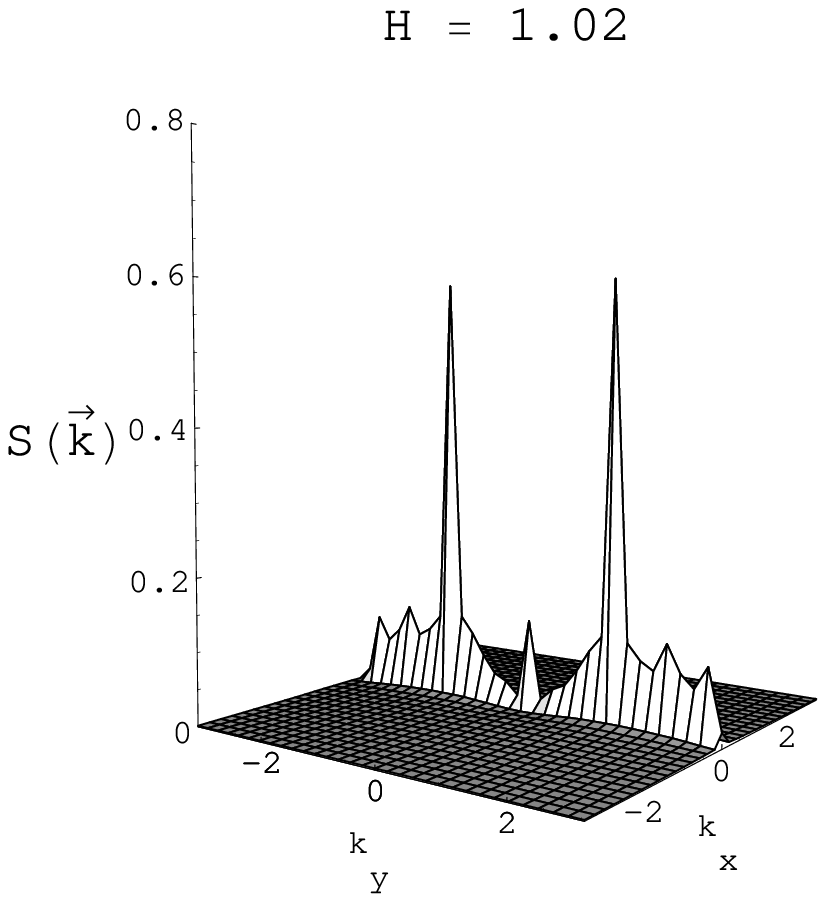}\\
\includegraphics[width=4.2cm,height=4.2cm]{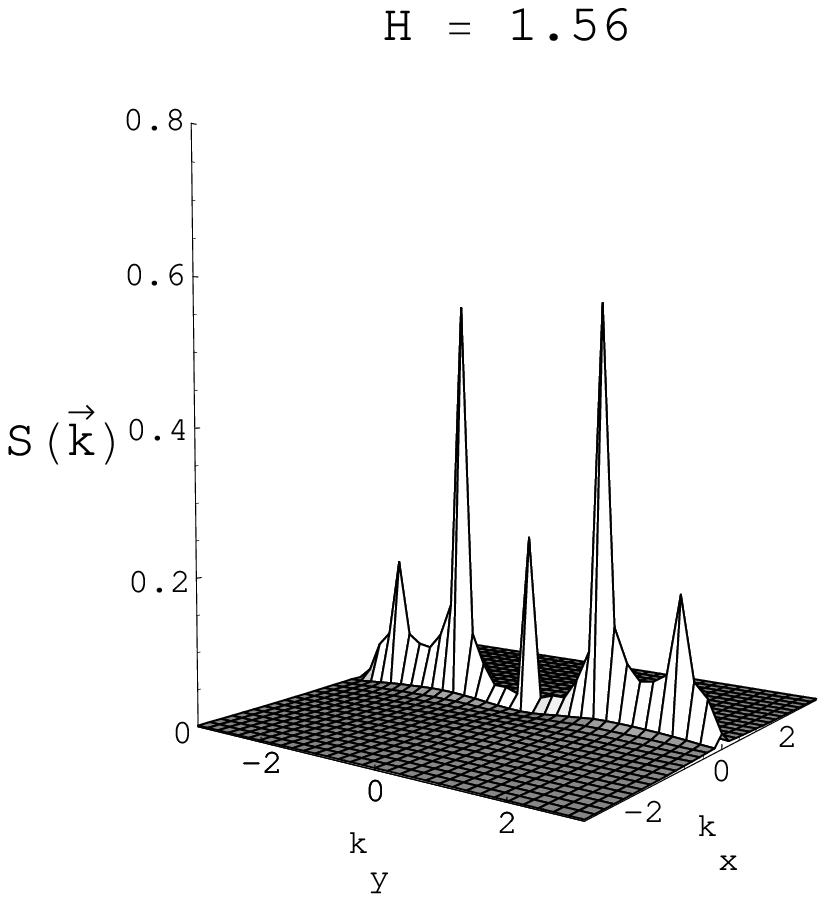}
\includegraphics[width=4.2cm,height=4.2cm]{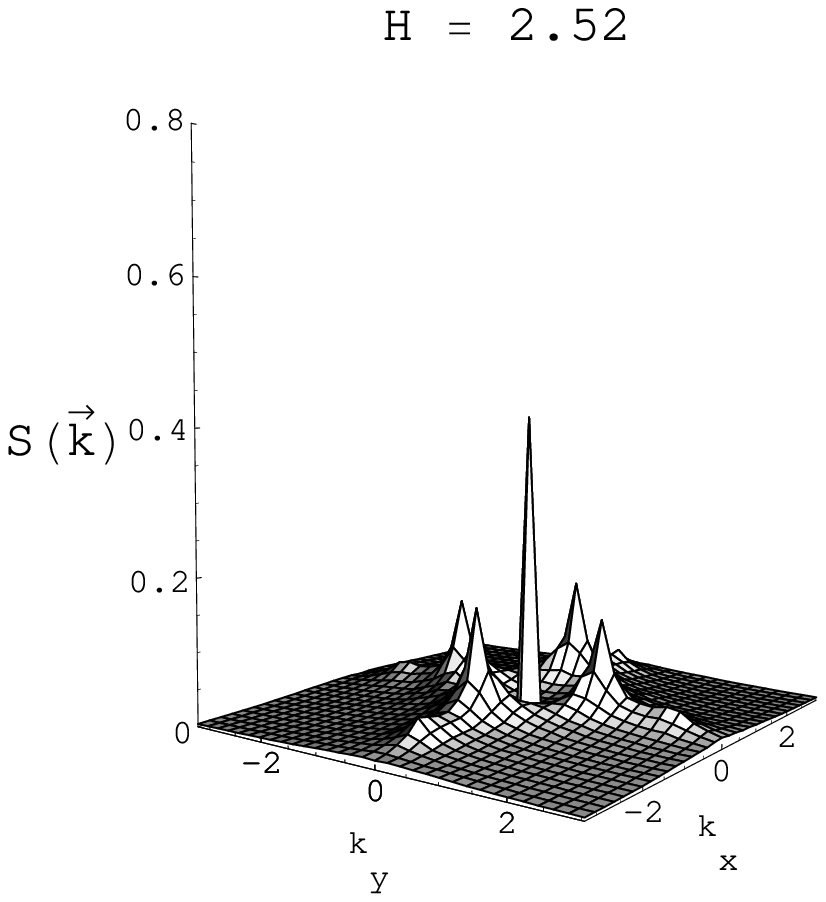}\\
\includegraphics[width=4.2cm,height=4.2cm]{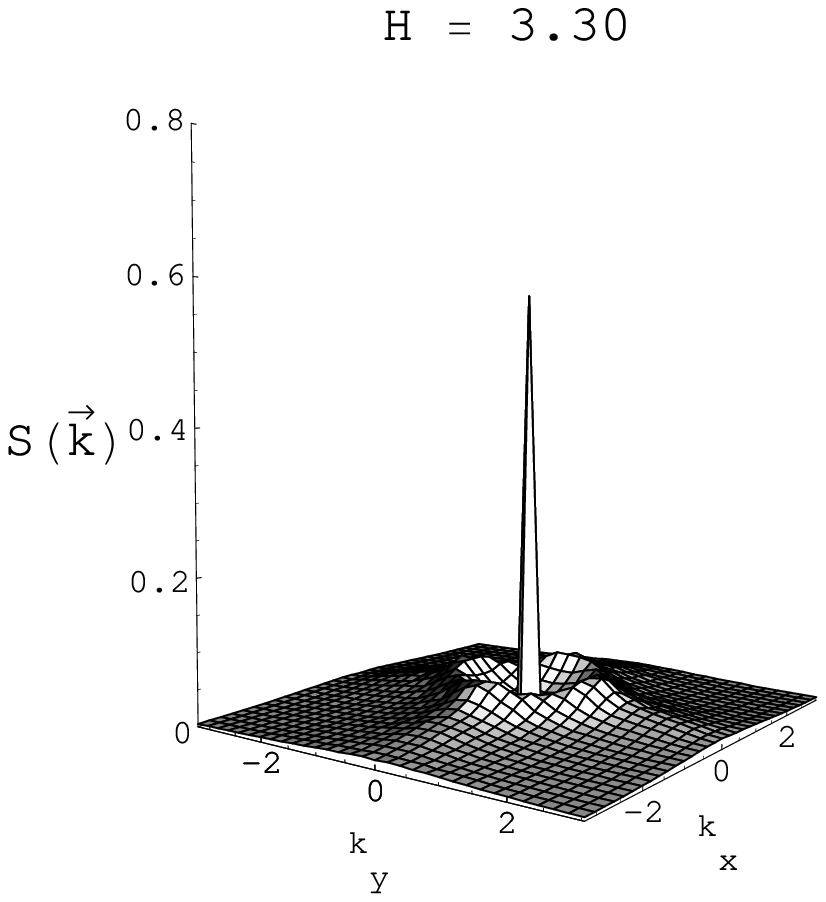}
\includegraphics[width=4.2cm,height=4.2cm]{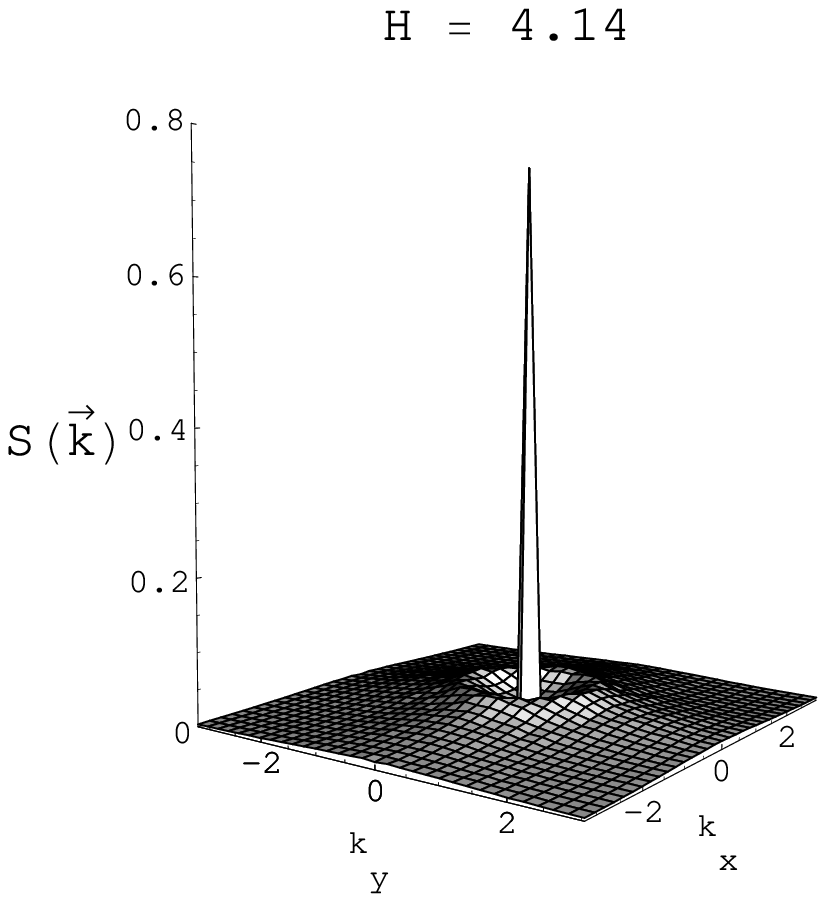}
\caption{Evolution of structure factor in a system of $L=32$ for increasing magnetic field.}
\label{sf3D}
\end{figure}

Figure \ref{sf3D} shows the structure factor of the system for different values of $H$. Each plot is obtained by the average of 5000 equilibrium configurations. At very low magnetic field, the system is characterized by a peak at one wave vector $\vec{k}=(0,\pi/2)$. Increasing $H$ new peaks appear  in $S(\vec{k})$.
First, with component $\vec{k} = (0, k_y \neq 0)$, still signaling the presence of orientational order in one direction. This change in the form of the structure factor is not evident a priory. One may, for instance, expect that the external magnetic field unbalances the number of up-down spins creating defects that  breaks the orientational order. Our results suggest a different scenario, where if properly equilibrated at low temperatures, new structures, without evident defects, keep the orientational long range order of the original ground state structures. Then, at higher magnetic fields, (see in the figure $H=2.52$) $S(k)$ becomes symmetric in both axis, the system looses the orientational order and reaches the bubble phase. Finally the magnetization saturates and only the peak at $\vec{k}=(0,0)$, survives.

\begin{figure}[!htb]
\includegraphics[width=7cm,height=7cm,angle=-90]{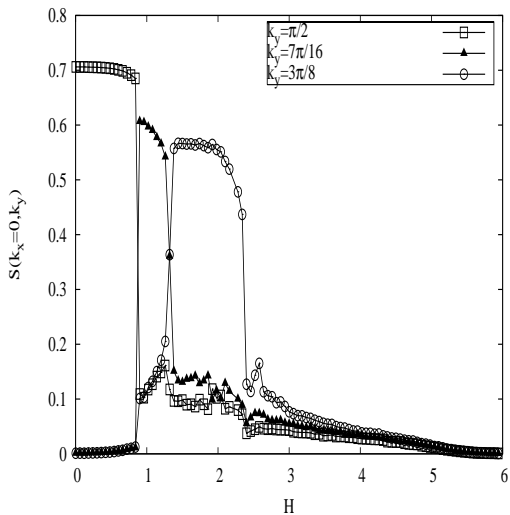}
\caption{Main harmonic contributions to the equilibrium configuration structure factor for increasing $H$ in a system of $L=32$. Other $k_y$ contributions remains always under $0.2$.}
\label{pky}
\end{figure}

Figure \ref{pky} shows the contribution of the three principal wave vectors $\vec{k}=(0,k_y^{*})$  characterizing the evolution of the system configurations with the magnetic field. 
Initially, the perfect stripes phase is characterized, as we know, by a wave vector $\vec{k}=(0,\pi/2)$. 
At $H \approx 0.84$ a new wave vector $\vec{k}=(0,7\pi/16)$ dominates the system, still indicating the presence of oriented stripes. 
Increasing further the magnetic field, at $H \approx 1.34$, $S(\vec{k})$ changes again, and $\vec{k}=(0,3\pi/8)$. The sudden rise and decay of each wave vectors reflects again the abrupt changes in the symmetry of the system.

\begin{figure}
\includegraphics[width=6.0cm,height=6.0cm,angle=-90]{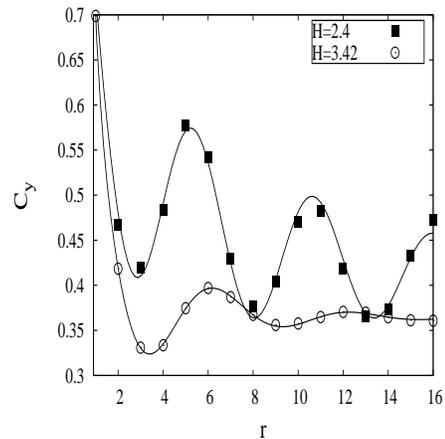}
\caption{Fits of correlation numerical data with the function (\ref{corr}); correlation best fits for two particular values of applied field}
\label{fcorrel}
\end{figure}

We also calculated  the directed spatial correlation functions for the system

\begin{eqnarray}
\nonumber
C_x(r)=\frac{1}{N}\sum_y\sum_x\langle S_{x,y}S_{x+r,y}\rangle\\
\nonumber
C_y(r)=\frac{1}{N}\sum_y\sum_x\langle S_{x,y}S_{x,y+r}\rangle\\
\nonumber
\end{eqnarray}
which reveal interesting information about the equilibrium states. In particular, we tried to fit the numerical data with a function of the form
\begin{equation}
C(r)=A e^{-\frac{r}{\xi}} \mathrm{cos}(k r-\psi)+B r^{-\alpha}+D
\label{corr}
\end{equation}

\noindent that has been proposed for the approximated continuum model\cite{mendoza,mulet07}. Figure \ref{fcorrel} shows the corresponding fits for averaged equilibrium configurations at two values of $H$.

\begin{figure}[!htb]
\includegraphics[width=6.0cm,height=10.0cm,angle=-90]{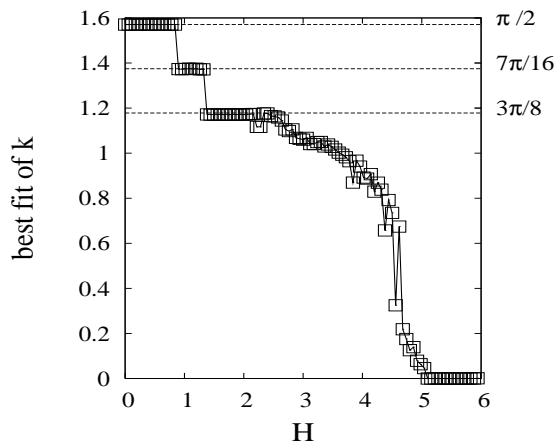}
\caption{Best fitted $k$ value as a function of $H$ in a system of $L=32$.}
\label{fcorrelb}
\end{figure}

From these fits we can gain information about the dependence with $H$ of the correlation length of the modulated domains ($\xi$), the main wave vector of the phase ($k$) and the power law strength ($\alpha$) respectively.
In particular, we can see in figure  \ref{fcorrelb} the behavior of  $k$ as a function of $H$. The plateaus in $k$ coincide with the principal wave vectors (see figure \ref{pky}) characterizing the different stripe structures.

To our knowledge these new anharmonic phases have  not been predicted before in a model with dipolar interactions. They are absent in the continuous model, where the effect of the magnetic field in the striped phase, is considered assuming that below the bubble phase the stripes persist in an increasing magnetized background \cite{garel82}.  They are present in the ANNNI model, but differently from \ref{ham}, the ANNNI model is anisotropic by construction.

On the other hand, in the phase diagram resulting from the simulations in \cite{arlett96} the orientational order parameter changes continuously from a finite value to zero at a given field (see figure 7 in that reference).  The reasons for these differences in the phase diagrams are not clear. We are tempted to think that looking at the dependence of $\eta$ for lower values of the temperature the authors in ref. \cite{arlett96} could find similar jumps and phases. Of course, having a large $\delta$ and hence larger stripe widths the anharmonicity properties of their structures may be hidden by strong finite size effects.

\subsection{Ground state analysis}

To study what kind of structures are responsible of the anharmonic phases, we tested the energy of a large number of configurations of alternating $S=-1$ and $S=1$ stripes. The width of the $S=-1$ stripes was varied from $1$ to $2$ while the width of $S=1$ stripes was varied from $0$ to $L$, as it is expected for the striped configurations in the presence of a field $H>0$.  Thus, borrowing the notation from ref. \cite{grous00} we denoted as $h24$ one configuration with stripes of width $2$ against the field and stripes of width $4$ in the field direction, repeated periodically. 

\begin{figure}
\includegraphics[width=6.0cm,height=6.0cm,angle=-90]{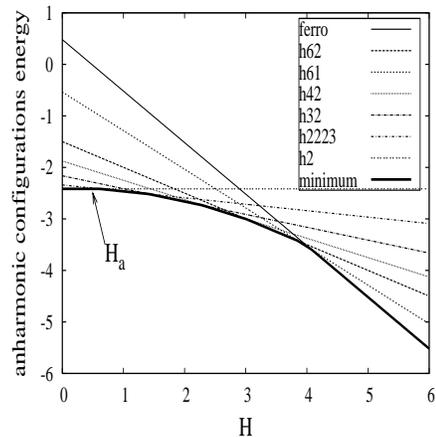}
\caption{Energy as function of $H$ in a system of $L=32$. Different anharmonic configurations at $T=0$ and the ground state energy (continuous line) are shown. $H_a$ is the critical field at which perfect stripes are lost. }
\label{energy}
\end{figure}

The energies of these configurations are represented in figure \ref{energy} as a function of $H$.
At $H=0$, the ground state of the system corresponds to the $h2$ phase. By increasing $H$ the system 
reaches a critical field $H_a$, where perfect stripes becomes energetically unfavorable with respect to the anharmonic configuration ($h2322$, $k_y=7\pi/16$). 
For larger fields, a new anharmonic configuration becomes the ground state ($h32$, $k_y=3\pi/8$). Further increasing $H$ the situation repeats with the appearance of new anharmonic states. How many of these anharmonic configurations may appear  depends strongly on temperature and commensuration effects. The corresponding ground state energies of the system, considering only these anharmonic configurations is represented in figure \ref{energy} with a continuous line. This line  corresponds to the lower energy 
curve obtained from the superposition of the energies of the different configurations as a function of $H$.

\begin{figure}
\includegraphics[width=6cm,height=8cm,angle=-90]{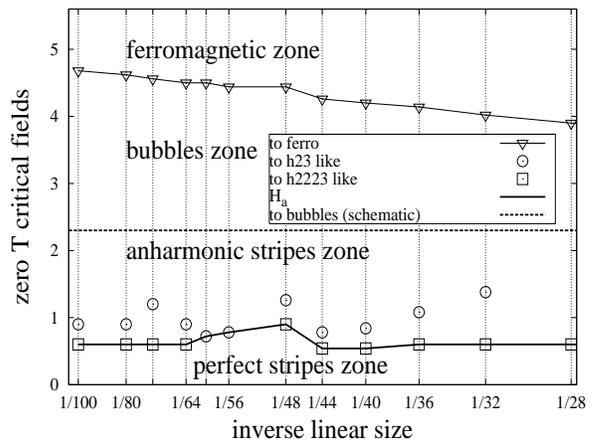}
\caption{Critical fields changing configurations at $T=0$ versus system size. $H_a$ and ferro lines are bolded for clarity but vertical lines were the only region explored and no interpolation is obvious. An schematic bubble line is also drawed (see the text).}
\label{criticalT0}
\end{figure}

One may wonder whether these are  finite size effects, and a non-orientatied ground state structure may dominate the behavior of the infinite system al low $H$. To test our predictions, this analysis was repeated for different system sizes, $N=L \times L$. Figure \ref{criticalT0} suggests that independently of the system size, the first critical field $H_a$ always appear in the low field region where the perfect stripes become unstable. 
This value defines a zone in which anharmonic structures establish, mainly in the form of $h2223$ or $h23$ configurations depending on commensuration effects.

The transition between anharmonic and bubbles phases remains around $H=2.4$ for system sizes up to $L=48$, this have been used to draw a schematic broken line in figure \ref{criticalT0}. 
Since the bubble phase establishes becouse entropic effects, this is lakely to be valid for large system sizes. 
In all tested cases the critical field remains well below this schematic transition, 
supporting the existence of the anharmonic phases obtained for $L=32$ in the thermodynamic limit, and
giving rise to a rather wide anharmonic zone.

\subsection{Phase Transitions}
\label{ssection:PT}

Unfortunately the computational cost associated with the presence of long range interactions and the commensuration effects in this kind of systems, prevent us from doing a proper finite size scaling analysis to define the character of the transitions. Instead, we focused our attention in systems of sizes  $L=32$ and $L=40$ and study
 the histograms of the energy and the order parameter.


\subsubsection*{Evidence for First Order Phase Transition}

\begin{figure}[!htb]
\includegraphics[width=5cm,height=4.2cm,angle=-90]{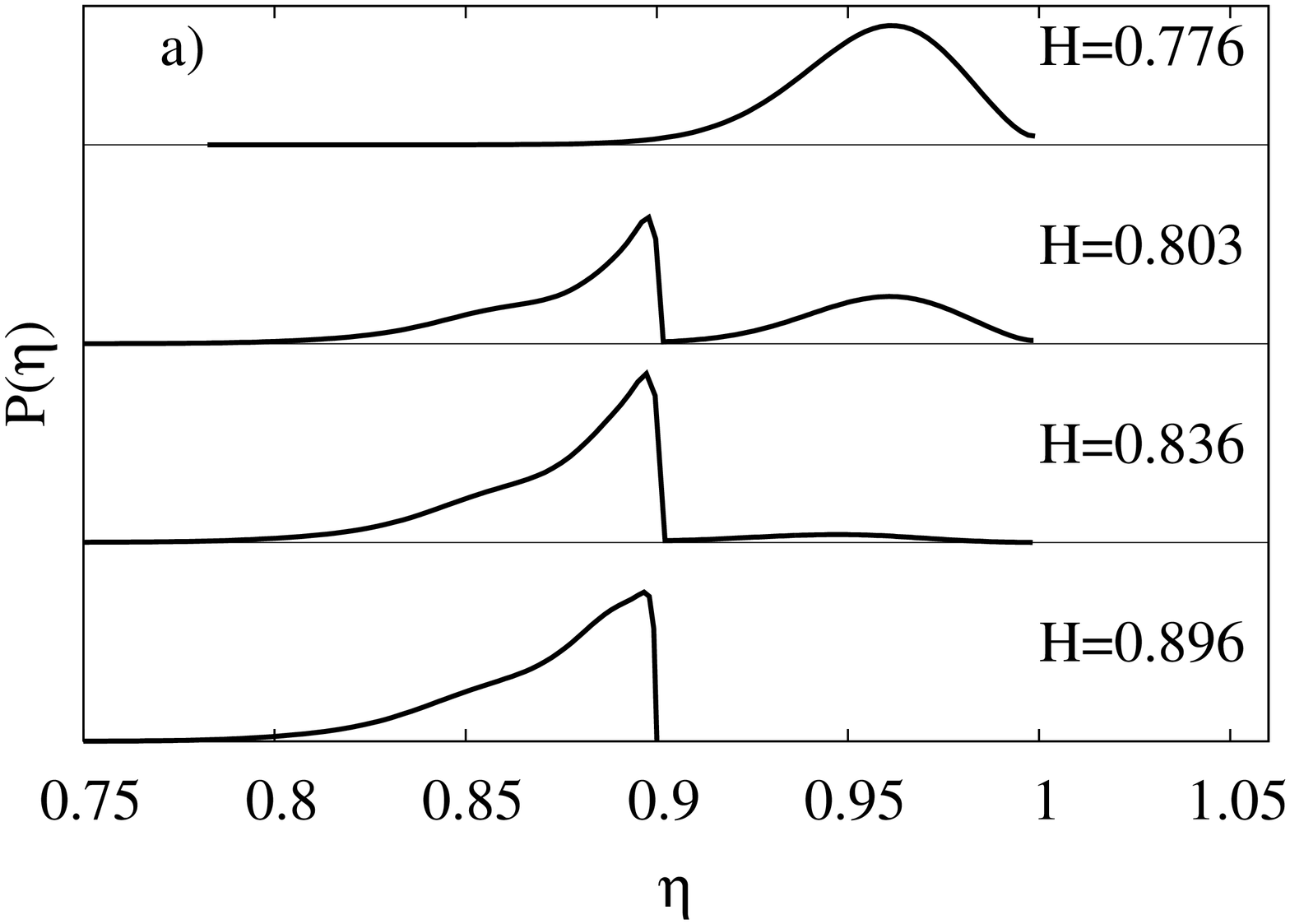}
\includegraphics[width=5cm,height=4.2cm,angle=-90]{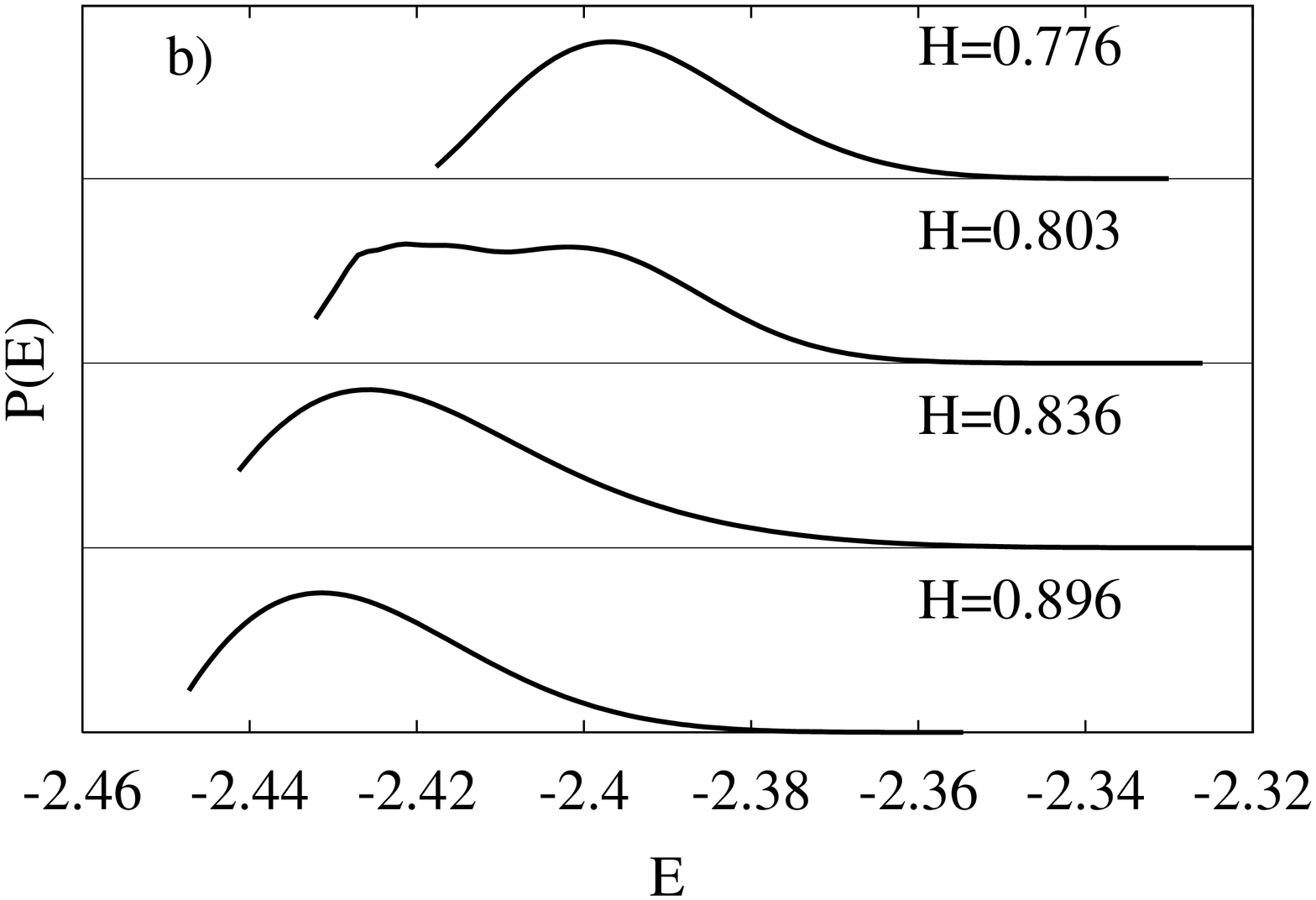}\\
\includegraphics[width=5cm,height=4.2cm,angle=-90]{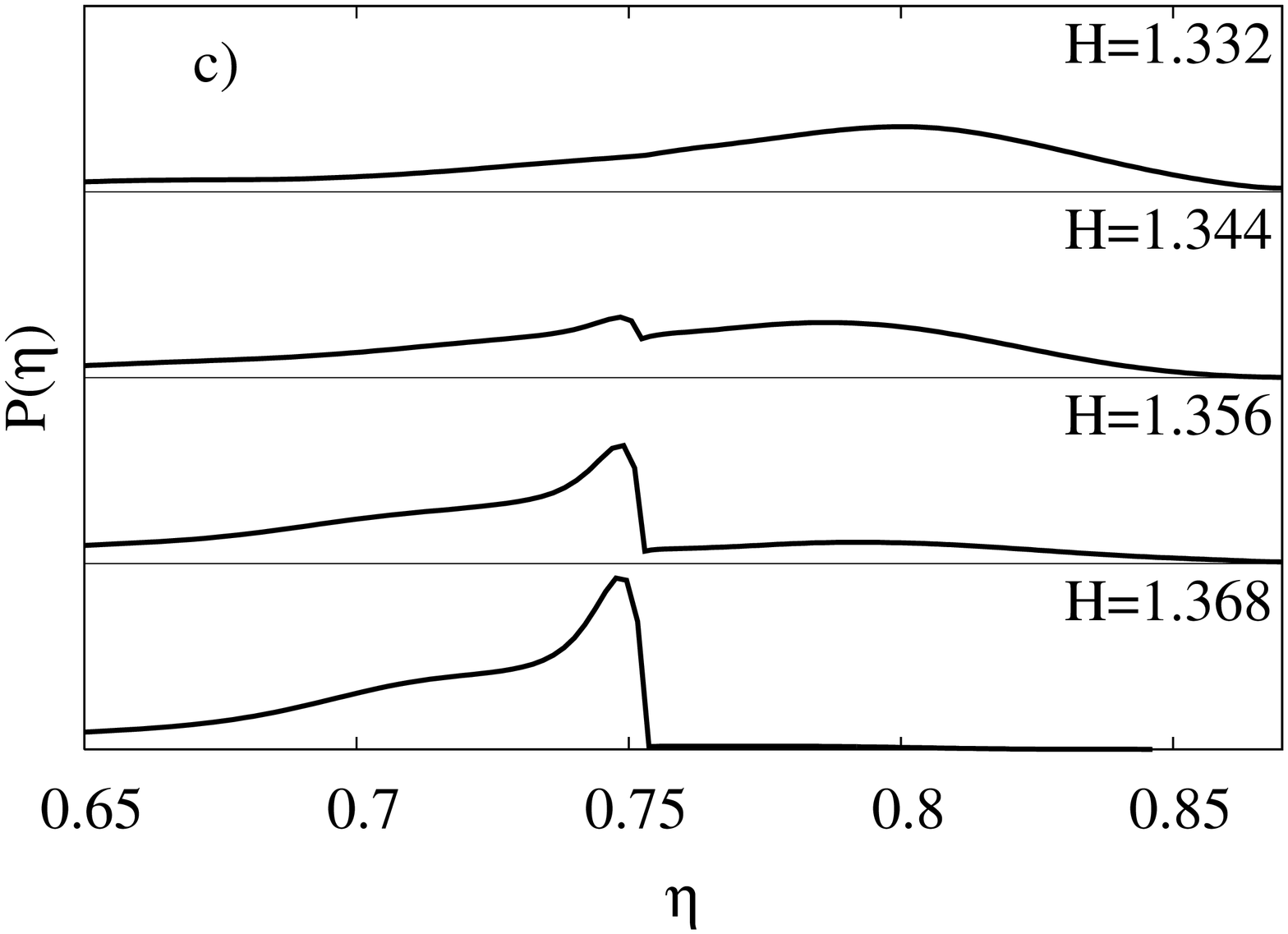}
\includegraphics[width=5cm,height=4.2cm,angle=-90]{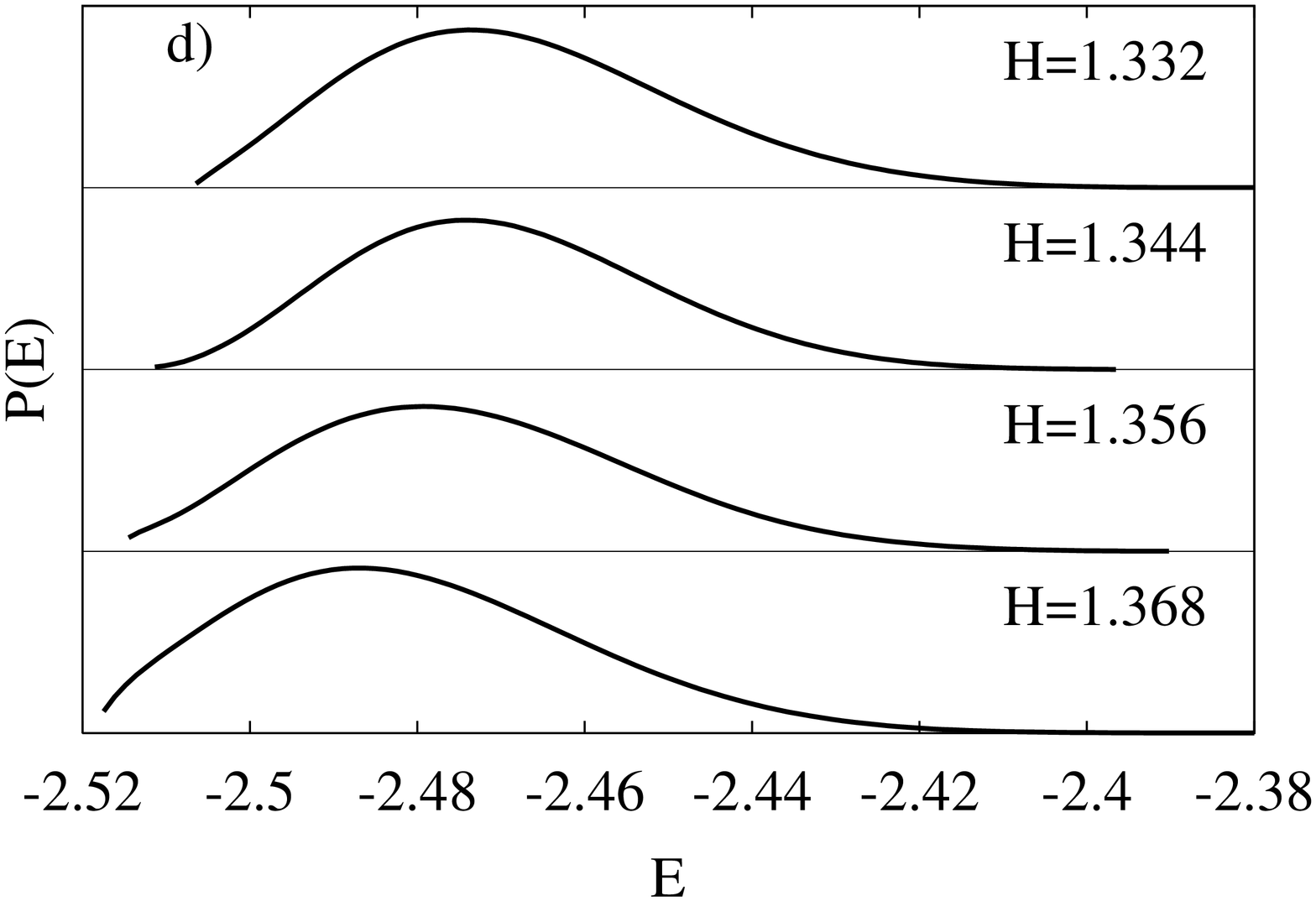}\\
\includegraphics[width=5cm,height=4.2cm,angle=-90]{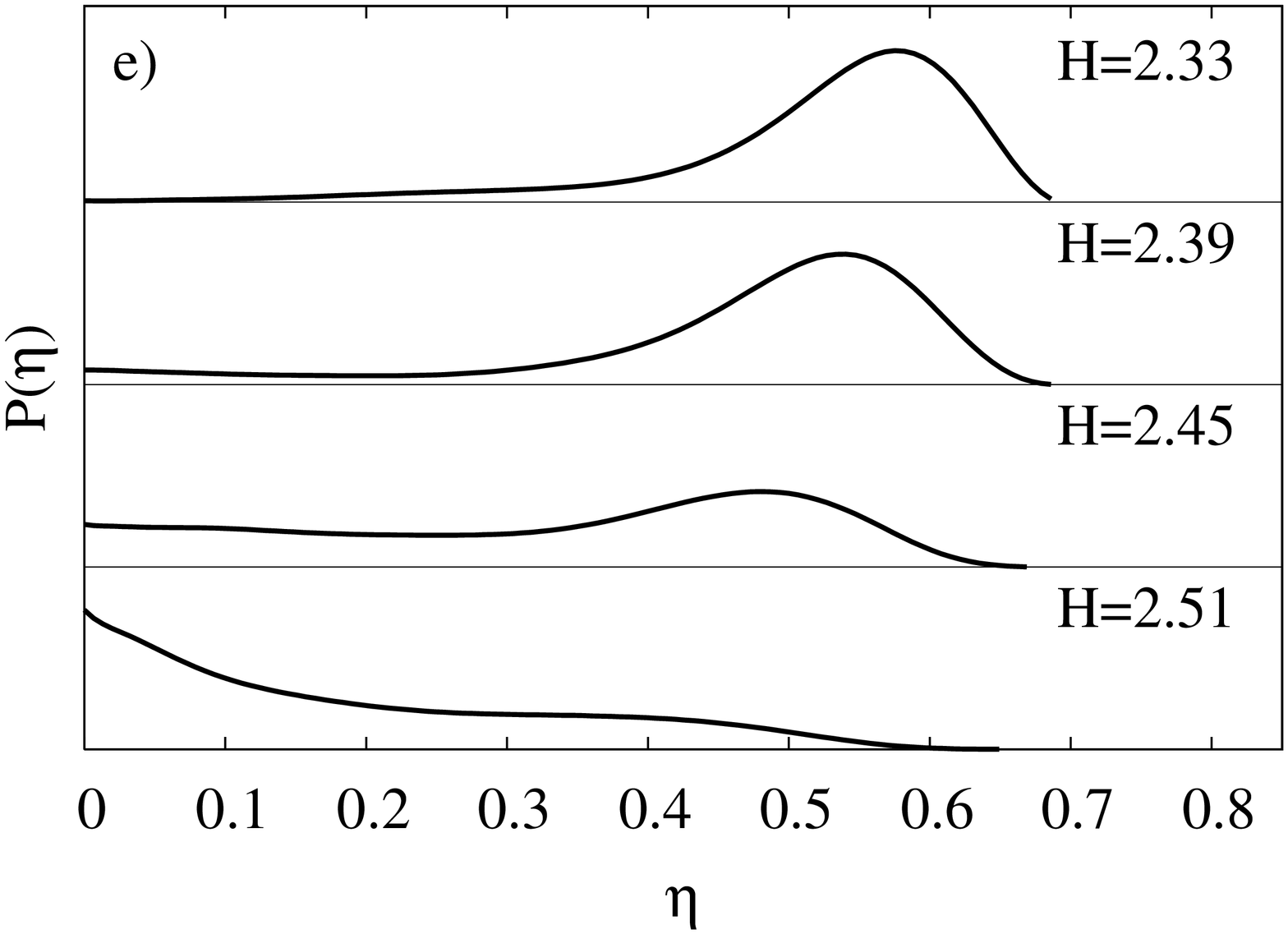}
\includegraphics[width=5cm,height=4.2cm,angle=-90]{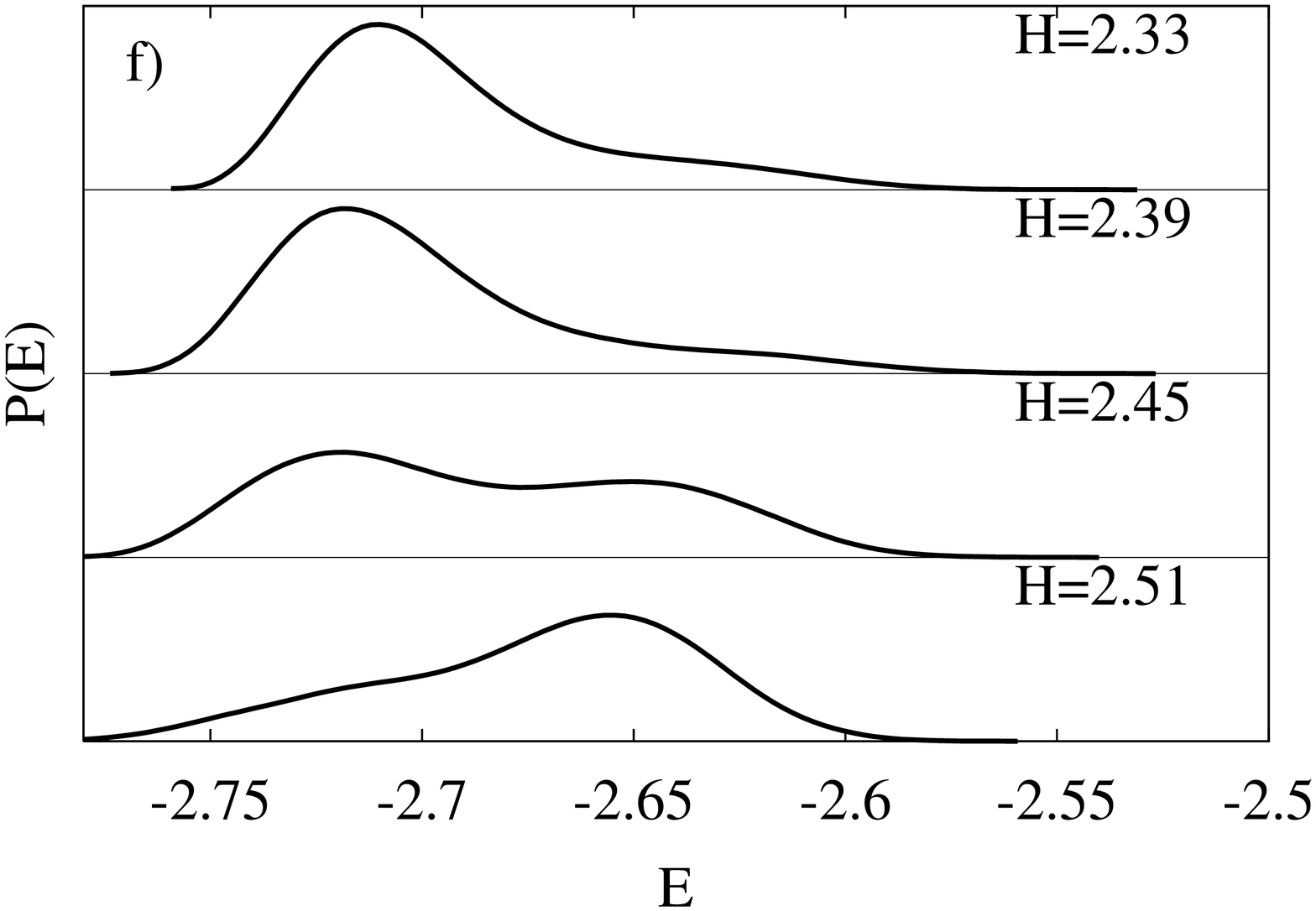}
\caption{Histograms of the orientational order parameter and energy for fields around the transitions involving striped phases: a) and b) for harmonic-anharmonic, c) and d) anharmonic-anharmonic, e) and f) anharmonic-bubbles.}
\label{s-a}
\end{figure}

The jumps in the susceptibilities and the discontinuities in $\eta$ in figure \ref{evo} already suggest the first order character of the transitions between the different orientational phases, and from the last anharmonic phase to the bubble phase. However, a stronger evidence is given in figure  \ref{s-a}. These histograms were calculated for systems of $N=40 \times 40$ spins, sampling $10^{7}$mcs after relaxation for each value of $H$ and considering $10^{5}$ values of energy and $\eta$.

For the three transitions considered, the figure shows that, increasing $H$,
the histograms of the order parameter and energy evolve from unimodal functions 
 at low magnetic fields, to a two peak shape structure, that disappears at higher magnetic fields giving rise to the new thermodynamic phase.  For the particular case of the anharmonic-anharmonic transition (c-d), the difference in energies between the two structures is so small that the histograms for the energy appear always as unimodal. 

On the other hand, one must note that while in the first two transitions, the peak in $P(E)$ moves from high to low energies, in the anharmonic to bubble transition it moves from low to high energies. In this transition, the system looses the orientational order and therefore $E$ increases.
 This is compensate by the presence of strong entropic effects that, in this more disordered structure, dominate the equilibrium state of the system.

\begin{figure}[!htb]
\includegraphics[width=6.0cm,height=6.0cm,angle=-90]{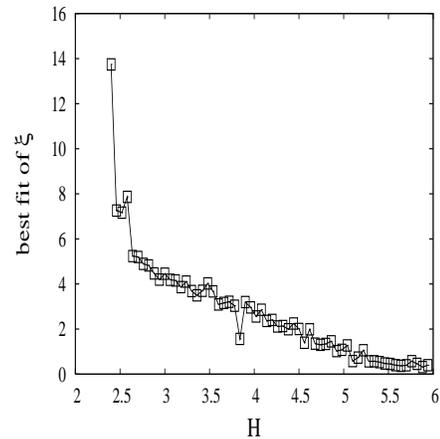}
\caption{Best fit of correlations by expression (\ref{corr}). Correlation length of modulated domains as a function of $H$ in a system of $L=32$, for $H$ below the bubbles region $\xi\geq L$.}
\label{chis}
\end{figure}

It is relevant for the definition of the anharmonic to bubble transition  the appearance at high fields of a non-zero correlation length  $\xi$ for the modulated domains (see figure \ref{chis}). Fitting the spatial correlations in the bubble phase with expression (\ref{corr}) we obtain the expected inverse proportionality of $\xi$ with  the applied magnetic field \cite{mendoza}. 

\subsubsection*{Evidence for a Kosterliz-Thouless transition}

\begin{figure}[!htb]
\includegraphics[width=8cm,height=2cm]{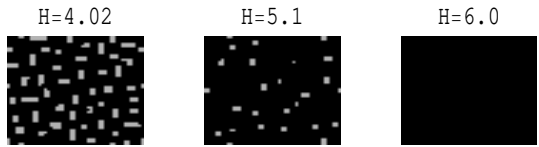}
\caption{Some field values involved in the transition from bubble to ferromagnetic phases. Typical configurations for an L=48 system.}
\label{pictures}
\end{figure}

\begin{figure}[!htb]
\includegraphics[width=5cm,height=8cm,angle=-90]{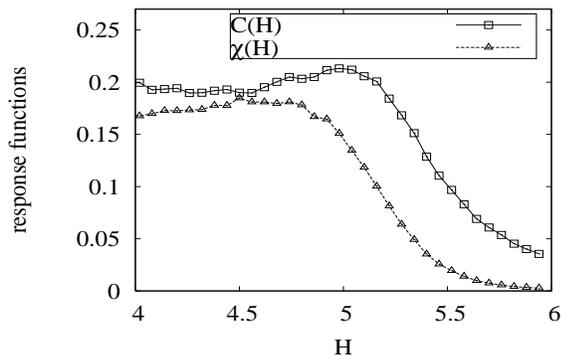}
\caption{Field values involved in the transition from bubble to ferromagnetic phases. Specific heat and magnetic susceptibility versus field in a system of $L=32$.}
\label{b-f}
\end{figure}

 Figure 
\ref{pictures} shows some views of the domain structure of the system close to the bubble-ferromagnetic transition. They  suggest that increasing $H$ the bubble phase dilutes in a ferromagnetic environment.  This support the predictions in 
\cite{garel82} where the authors proved that within a Ginzburg-Landau approximation, dislocation of the bubbles structure may lead to a second-order melting transition of the Kosterliz-Thouless type.  The continuous change of energy and magnetization (see figure \ref{evo}a) and the saturation of the response functions close to this transition (see in figure \ref{b-f} zooms of the magnetic susceptibility and the specific heat close to this transition) also support these predictions.

\begin{figure}
\includegraphics[width=6.0cm,height=6.0cm,angle=-90]{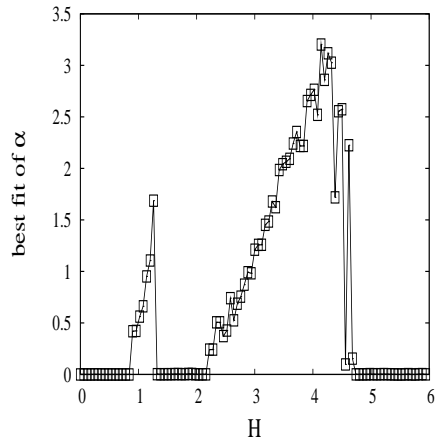}
\caption{Best fit of correlations by expression (\ref{corr}). Power exponent  as a function of $H$ in a system of $L=32$.}
\label{chisb}
\end{figure}

One last indication in favor of this scenario, comes from the spatial correlation functions of the system.  In figure \ref{chisb} we show the value of $\alpha$ obtained by the fits of the correlation functions with expression (\ref{corr}). The sudden rise of $\alpha$ close to $H \sim 4.5$ is also consistent with a Kosterliz-Thouless transition.

\section{Conclusions}
\label{conc}

We developed extensive numerical simulations to characterized the phase diagram of the model given by (\ref{ham}). This Hamiltonian presents at very low field and temperature a phase of symmetric stripes and zero magnetization. Increasing the field, new thermodynamical phases appear, still with orientational order but with non-zero magnetization and characterized by different wave vectors. As far as we know, the existence of these thermodynamic phases have not being proposed before for systems with dipolar interactions. For larger values of $H$, the system enters into the bubble phase loosing the orientational order. Then, at larger fields, the system becomes fully magnetized.

We present evidence supporting the idea that all, but the bubble to ferromagnetic, are first order transitions. This is also in agreement with analytical results that predicted that the stripes to bubbles transition is of the Brazovskii \cite{bra75} type. On the other hand, close to the bubbles to ferro transition, our simulations show the existence of a continuous order parameter, the saturation of the response functions and algebraically decaying spatial correlations, supporting all, a Kosterliz-Thouless scenario. 

Finally, it is worth to note the interesting parallelism between these anharmonic phases and the hybrid states found for Hamiltonian (\ref{ham}) at zero field in  reference \cite{pig06}. There, through Mean Field calculations, the authors suggested a possible interpretation of nematic phases as a competition between striped structures of different widths.
Moreover, they found Kosterliz-Thouless features in the transition between striped and nematic phases. To clarify these issues, and to completely define the phase diagram (\ref{diag}) more accurate simulations are expected close to the $H=0$  critical temperature.


\begin{acknowledgments}
We gratefully acknowledge partial financial support from the Abdus Salam ICTP through grant {\em Net-61,
La\-tin\-a\-me\-ri\-can Network on Slow Dynamics in Complex Systems}. We thank D. Stariolo and S. Cannas for useful comments on a previous manuscript. Calculation facilities kindly offered by the Bioinformatic's Group of the Center of Molecular Inmunology in Cuba were instrumental to this work.
\end{acknowledgments}

\bibliography{isingH_v1.7}

\end{document}